\documentclass[a4paper,fleqn]{cas-dc}

\usepackage[authoryear]{natbib}
\usepackage{amsmath}
\usepackage{booktabs}
\usepackage{graphicx}
\usepackage{array}
\usepackage{float}
\usepackage{caption}

\begin{document}
\let\WriteBookmarks\relax

\shorttitle{Import-Dependent Grain Processing Hubs}
\shortauthors{M.L. Kurnaz}

\title[mode=title]{Import-Dependent Grain Processing Hubs: The Case of T\"{u}rkiye's Flour Sector}

\author[1,2]{M. Levent Kurnaz}
\cormark[1]
\ead{levent.kurnaz@bogazici.edu.tr}

\credit{Conceptualization, Methodology, Formal analysis, Writing -- Original Draft, Writing -- Review \& Editing}

\affiliation[1]{organization={Bo\u{g}azi\c{c}i University, Center for Applied Research in Finance},
                addressline={Bebek},
                city={Istanbul},
                country={T\"{u}rkiye}}

\affiliation[2]{organization={Bo\u{g}azi\c{c}i University, Center for Climate Change and Policy Studies},
                addressline={Bebek},
                city={Istanbul},
                country={T\"{u}rkiye}}

\cortext[1]{Corresponding author}

\begin{abstract}
International commerce has long been seen as a key way to keep the global food system stable,
allowing agricultural surpluses in some areas to compensate for shortages in others. This
strategy has led to the rise of highly specialised processing hubs that combine significant
industrial capacity with agricultural inputs sourced from throughout the world. T\"{u}rkiye's
flour sector---currently the largest wheat flour exporter in the world---represents one of the
most prominent examples of this model. However, increasing climate variability and geopolitical
fragmentation raise important questions regarding the long-term resilience of food systems that
rely heavily on imported biological inputs. Recent research shows the growing probability of
synchronised crop failures across multiple agricultural regions due to atmospheric circulation
anomalies and climate-induced extreme weather events. The assumption that global markets can
consistently rebalance supply disruptions through trade is challenged by such events. Using the
flour industry of T\"{u}rkiye as a case study, this paper investigates the susceptibility of
globally integrated grain processing centres. In order to assess the correlation between the
scope of industrial processing and the capacity of domestic agricultural production, we introduce
the Biophysical Autonomy Ratio~(BAR). The analysis demonstrates that T\"{u}rkiye's BAR has
declined consistently over time, suggesting that its processing sector has expanded beyond the
domestic production base. The results suggest that in order to enhance the resilience of the
food system in the future, it may be necessary to establish a more precise alignment between
biological production systems and industrial food infrastructure. The paper concludes by
addressing the policy implications for national food security governance in the context of
escalating climate instability.
\end{abstract}


\maketitle

\section{Introduction}
\label{sec:intro}

Over the past 50 years, the world's food system has become a more interconnected network of
production areas, processing industries, and trade routes. As climate change worsens and global
agricultural supply chains become more connected, policymakers are increasingly concerned about
the resilience of food systems that depend on international trade for important biological inputs.
International trade has been a significant part of this change, allowing countries to use global
markets to compensate for gaps in regional production while focusing on areas where they have a
comparative advantage \citep{Headey2011, Puma2015}.

Because of this reliance on trade, specialised industrial nodes have been able to grow within
global food value chains. Countries that import large volumes of agricultural raw materials and
export processed food products to global markets often host large food processing sectors. These
processing hubs function as logistical and industrial intermediaries, transforming biological
inputs from around the world into value-added products for international distribution.

Although this model has enhanced efficiency and increased food availability, there is a growing
debate about its long-term resilience in the face of environmental change. Climate change is
expected to increase the frequency and severity of extreme weather events affecting agricultural
production, including heat waves, droughts, floods, and compound climate anomalies. Recent
research has highlighted the possibility of synchronised crop failures across multiple major
agricultural regions, referred to as Multiple Breadbasket Failure~(MBBF)
\citep{Anderson2019, Gaupp2019, Gaupp2020, Mehrabi2019, Tigchelaar2018, Hasegawa2022, Qi2022}.
Such events may occur when climatic anomalies or teleconnected atmospheric patterns affect
multiple agricultural regions simultaneously, generating correlated production shocks across
geographically distant food-producing areas.

These developments raise important questions about the resilience of food systems that depend
heavily on international agricultural supply chains. In particular, if global grain flows become
disrupted, countries that host large food processing industries but rely on imported biological
inputs may face new forms of systemic vulnerability.

T\"{u}rkiye's flour sector provides a particularly relevant case study. T\"{u}rkiye hosts one of
the world's largest flour-processing sectors while simultaneously relying partly on imported wheat
inputs, making it representative of a growing class of globally integrated food-processing hubs.
Over the past decade, T\"{u}rkiye has consistently been the world's largest exporter of wheat
flour. The sector's success has been made possible by a policy framework that allows the
tariff-free import of wheat for processing and re-export. This structure has enabled the country
to develop a large and technologically advanced milling industry capable of serving markets across
the Middle East, Africa, and Asia.

However, the scale of T\"{u}rkiye's milling capacity significantly exceeds domestic wheat
production. As a result, the sector relies heavily on imported grain supplies.

This paper examines the vulnerability of trade-dependent food processing hubs under conditions of
increasing climate instability. Specifically, it asks how countries whose food processing sectors
rely heavily on imported agricultural inputs may be exposed to systemic risks when climate-driven
disruptions affect global grain supply. Using T\"{u}rkiye's flour sector as a case study, the
paper introduces the concept of the Biophysical Autonomy Ratio~(BAR) as an analytical indicator
measuring the relationship between domestic biological production capacity and industrial
processing scale. By applying this framework to T\"{u}rkiye's flour industry, the study
highlights how structural imbalances between domestic production and processing capacity may
increase exposure to supply shocks in international grain markets. The paper contributes to
ongoing debates on food system resilience by proposing a policy-relevant diagnostic tool for
assessing trade-dependent food processing systems in an era of increasing climate volatility.

\section{Climate change and instability in global grain systems}
\label{sec:climate}

Global agricultural production has historically demonstrated a high degree of resilience through
geographic diversification. When crop failures occurred in one region, production surpluses
elsewhere could often compensate through trade. This mechanism has been a fundamental stabilising
feature of global food systems.

However, emerging research suggests that climate change may increase the probability of
synchronised agricultural disruptions across multiple major production regions. Several mechanisms
contribute to this possibility.

First, rising global temperatures increase the frequency of extreme heat events during critical
crop development periods. Wheat, maize, and other staple crops are particularly sensitive to heat
stress during flowering and grain-filling stages \citep{Asseng2014, Zhao2017, Schauberger2017,
Lobell2011}. Extreme temperatures during these periods can significantly reduce yields, and
empirical analyses of disaster records confirm that droughts and heat events have caused
nationally significant cereal production losses \citep{Lesk2016}.

Second, climate change influences atmospheric circulation patterns that shape regional weather
variability. Studies of atmospheric dynamics suggest that the disproportionate warming of the
Arctic relative to lower latitudes (Arctic amplification) may influence the behaviour of
mid-latitude jet streams \citep{Francis2012}. These changes can amplify Rossby wave patterns,
potentially leading to persistent weather extremes such as prolonged droughts or heat waves
across multiple regions simultaneously \citep{Kornhuber2020}.

Third, climate change may intensify compound events in which multiple stressors occur
simultaneously or sequentially \citep{Raymond2020, Biess2024, Chatzopoulos2021}. For example,
drought conditions combined with extreme heat can severely reduce crop yields, while flooding
during planting seasons may prevent cultivation altogether.

These mechanisms raise the possibility of multiple breadbasket failures, in which several major
agricultural exporting regions experience significant yield reductions during the same growing
season \citep{Gaupp2019, Gaupp2020, Hasegawa2022, Tigchelaar2018}. Such events could reduce
the capacity of global markets to compensate for regional production shocks
\citep{Anderson2019, Heino2020}.

Recent historical episodes illustrate the potential consequences of such disruptions. Production
shocks in several key exporting regions influenced the global food price crises of 2007--2008
and 2010--2011 \citep{Headey2011}. During these crises, many countries implemented export
restrictions in an attempt to protect domestic consumers. These policy responses reduced the
availability of grain on international markets and contributed to further price volatility.

More recently, geopolitical disruptions affecting Black Sea grain exports demonstrated how
concentrated supply chains can amplify systemic risk. When a limited number of regions account
for a large share of global exports, disruptions affecting those regions can have cascading
effects across international food systems \citep{Keys2025, Caparas2021}.

Under conditions of increasing climate variability, the stability of trade-based food security
strategies may therefore become more uncertain. Countries that rely heavily on global markets for
staple food supplies may face greater exposure to both environmental and geopolitical disruptions.

\section{Globally integrated food processing hubs}
\label{sec:hubs}

Within the global food system, certain countries have developed large food processing industries
that function as intermediaries between agricultural production regions and consumer markets.
These processing hubs import agricultural raw materials, transform them into processed products,
and export them to international markets \citep{Puma2015, BrenDAmour2016, Cottrell2019}.

Such industries benefit from economies of scale, advanced infrastructure, and access to global
trade networks. However, they may also depend heavily on continuous access to imported biological
inputs.

Often, these sectors operate under policy frameworks designed to support export-oriented industrial
production. These frameworks frequently allow tariff-free import of agricultural inputs that are
processed and subsequently re-exported as value-added products. While these policies enhance
industrial competitiveness, they may also create structural dependence on global supply chains.

This configuration creates a particular form of systemic vulnerability when industrial processing
capacity becomes significantly larger than domestic agricultural production capacity
\citep{Gaupp2020, Keys2025, Deteix2024}. In such cases, the functioning of the processing
sector depends on the continuous availability of imported biological inputs.

This model can produce substantial economic benefits when international markets function smoothly.
However, under conditions of global supply disruption, processing hubs may experience input
shortages that constrain industrial output.

Understanding this relationship between domestic agricultural production and industrial processing
scale is therefore important for evaluating the resilience of food systems under conditions of
environmental change.

\section{Analytical framework: The Biophysical Autonomy Ratio}
\label{sec:bar}

To evaluate the structural relationship between domestic agricultural production and
export-oriented food processing, this study introduces the Biophysical Autonomy Ratio~(BAR) as
a simple diagnostic indicator.

The BAR measures the extent to which a country's domestic biological production capacity is
sufficient to support both internal consumption and the demands of its processing sector. In
contrast to conventional food security indicators, which typically focus on domestic supply
relative to domestic demand, the BAR explicitly incorporates the additional demand generated by
participation in global value chains.

Formally, the BAR is defined as:

\begin{equation}
\mathrm{BAR}_{i,t} = \frac{P_{i,t}}{A_{i,t} + E^{w}_{i,t}}
\label{eq:bar}
\end{equation}

\noindent where $P_{i,t}$ denotes domestic wheat production in country~$i$ at time~$t$,
$A_{i,t}$ represents apparent domestic wheat availability, and $E^{w}_{i,t}$ corresponds to
wheat-equivalent export demand generated by processed wheat products. Values of the ratio provide
a structural interpretation of system balance. A BAR value equal to or greater than one indicates
that domestic production is sufficient to support both consumption and processing activity, while
values below one imply reliance on imported biological inputs to sustain the current scale of
industrial activity.

The BAR is conceptually related to, but distinct from, existing metrics used to evaluate resource
flows in global food systems. Virtual water accounting \citep{Allan1998, Hoekstra2005} tracks the
volume of water embodied in traded agricultural commodities and has been widely applied to assess
hidden resource dependencies in food trade. The BAR addresses an analogous concern at the level
of agricultural production capacity rather than water use, asking how much of a country's
industrial food processing activity is supported by its domestic land and production base as
opposed to imported biological inputs. Similarly, virtual land transfer analysis provides a
related but production-side perspective on embodied resource dependencies by examining the land
resources implicitly traded through agricultural commodity flows \citep{Fader2011, Rulli2013}.
The BAR differs from these approaches in focusing specifically on the structural relationship
between export-oriented processing scale and domestic production capacity, rather than on the
aggregate volume of embodied resources in trade flows. It is also more directly
operationalisable from standard FAOSTAT data, which may make it useful as a practical monitoring
indicator in policy contexts. The indicator is not intended to replace these established
frameworks but to complement them by drawing attention to a dimension of food system structure
---namely the industrial load imposed on domestic agricultural systems by export-oriented
processing---that standard self-sufficiency ratios do not capture \citep{Clapp2017, Deteix2024}.

The conceptual contribution of the BAR lies in its ability to capture the industrial load imposed
on domestic agricultural systems. Standard self-sufficiency ratios compare production with
domestic consumption and therefore do not reflect the additional demand created by
export-oriented processing sectors. As a result, countries with large processing industries may
appear self-sufficient under conventional metrics while in practice relying heavily on imported
inputs. By incorporating export-related demand, the BAR provides a more complete representation
of the relationship between agricultural production systems and industrial food processing
capacity.

From a systems perspective, the BAR can be interpreted as an indicator of biophysical autonomy.
Higher values indicate that domestic ecosystems largely support industrial food production, while
lower values reflect an increasing dependence on external agricultural systems. Importantly, low
BAR values do not imply inefficiency or immediate instability; rather, they indicate structural
exposure to external supply conditions. Under stable global trade regimes, such configurations
can be economically advantageous. However, under conditions of synchronised production shocks or
trade disruptions, systems with low BAR values may face heightened vulnerability due to their
reliance on imported biological inputs.

\subsection{Operationalisation of the BAR indicator}
\label{subsec:operationalisation}

For the empirical analysis, the BAR is calculated using publicly available production and trade
statistics. Annual wheat production data were obtained from FAOSTAT. Apparent domestic wheat
availability was approximated by adding production and imports and subtracting exports, which
provides consistent estimates of wheat use over time.

Export volumes were converted into wheat-equivalent quantities because a significant portion of
wheat demand in processing hubs is embodied in exported products. The analysis focuses on the
dominant processed wheat products in international trade, namely wheat flour and pasta. A
standard industrial extraction coefficient of 0.75 was applied, implying that one tonne of wheat
yields approximately 0.75 tonnes of processed product. Wheat-equivalent export demand was
therefore calculated as:

\begin{equation}
E^{w}_{i,t} = \frac{F_{i,t}}{0.75} + \frac{\mathrm{Pasta}_{i,t}}{0.75}
\label{eq:wheat_equiv}
\end{equation}

\noindent where $F_{i,t}$ and $\mathrm{Pasta}_{i,t}$ denote flour and pasta export volumes,
respectively. All quantities were expressed in million tonnes~(Mt) to ensure comparability
across datasets.

To reduce interannual variability driven by climatic fluctuations, the resulting BAR time series
are presented as five-year moving averages. This smoothing allows the analysis to focus on
structural trends rather than short-term production shocks.

\subsection{Methodological considerations}
\label{subsec:methodology}

The BAR provides a simplified representation of complex food system dynamics and should therefore
be interpreted as a structural diagnostic rather than a precise accounting framework. Several
limitations should be noted.

First, the indicator does not explicitly account for stock changes, feed use, or variations in
industrial capacity utilisation, all of which may influence short-term supply-demand balances.
Second, the use of a standard extraction coefficient introduces a degree of approximation, as
conversion ratios may vary across processing technologies and product types. However, sensitivity
analyses using plausible alternative extraction coefficients do not materially affect the
observed trends.

Finally, the BAR does not capture adaptive responses such as substitution between suppliers,
changes in trade routes, or policy interventions, including strategic grain reserves. Despite
these limitations, the indicator offers a transparent and reproducible approximation of the
relationship between domestic biological production capacity and the scale of export-oriented
food processing.

As such, the BAR is not a measure of food security per se, but of the structural alignment
between industrial food systems and the ecological production base that sustains them.

\section{T\"{u}rkiye's wheat and flour sector}
\label{sec:turkiye}

T\"{u}rkiye's flour industry operates as a globally integrated processing hub at a scale that
frequently exceeds the domestic wheat production base. Over the past decade, the country has
consistently been the largest exporter of wheat flour worldwide. This success is supported by a
large and technologically advanced milling sector with substantial processing capacity, modern
logistics infrastructure, and geographic proximity to major wheat-producing regions and consumer
markets across the Middle East, North Africa, and Asia.

An important institutional factor supporting the sector is the Domestic Processing Regime, which
allows the tariff-free import of wheat for processing and subsequent re-export. Milling companies
can import grain from international markets without facing import duties under this framework, as
long as they export the processed products \citep{Headey2011}. This policy structure has
significantly enhanced the international competitiveness of T\"{u}rkiye's flour industry.

At the same time, the regime creates a structural separation between the scale of industrial
processing capacity and the country's domestic agricultural production base. While T\"{u}rkiye
is itself a major wheat producer, national production levels are generally lower than the total
wheat volume processed by the milling industry.

Long-term production statistics illustrate both the scale and the variability of T\"{u}rkiye's
domestic wheat supply. Over the past two decades, national wheat production has fluctuated
between approximately 17 and 22 million tonnes annually, largely reflecting climatic conditions
affecting rain-fed wheat regions in Anatolia. Although the long-term production base remains
relatively stable around 20 million tonnes, interannual variability of several million tonnes is
common due to drought and temperature stress during the growing season.

Figure~\ref{fig:Fig_1} compares this production trajectory with the estimated wheat
demand generated by domestic consumption and export-oriented processing activities. The
comparison highlights that the scale of wheat processing activity frequently exceeds the domestic
production base. This relationship implies that a portion of the milling sector's activity
depends on imported grain supplies to sustain current export volumes.

\begin{center}
  \includegraphics[width=\linewidth]{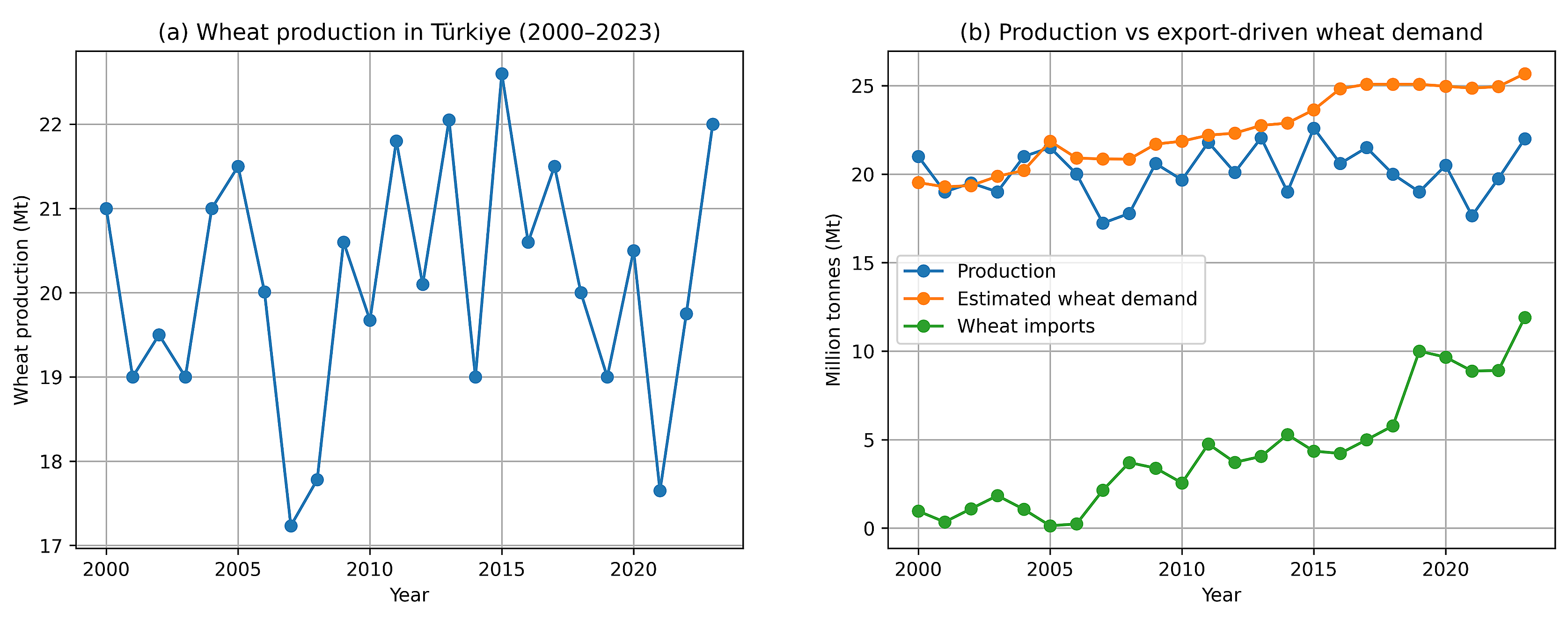}
  \captionof{figure}{Structure of T\"{u}rkiye's wheat system.
  (a)~Annual wheat production in T\"{u}rkiye between 2000 and 2023. Production fluctuates
  between approximately 17 and 22 million tonnes, depending largely on climatic conditions
  affecting rain-fed wheat regions.
  (b)~Domestic wheat production compared with estimated wheat demand generated by domestic
  consumption and wheat-equivalent volumes required to produce exported flour and pasta.
  Wheat-equivalent export demand is calculated using a standard flour extraction ratio of 0.75.
  The figure illustrates that the scale of T\"{u}rkiye's export-oriented wheat-processing sector
  frequently exceeds the domestic production base. Data sources: FAOSTAT trade statistics and
  national production estimates.}
  \label{fig:Fig_1}
\end{center}

The scale of export-oriented processing further illustrates this structural relationship.
T\"{u}rkiye exports approximately 3--4 million tonnes of wheat flour annually, corresponding to
roughly 4--5 million tonnes of wheat equivalent when standard milling extraction ratios are
applied. When combined with domestic consumption, the total wheat volume required to support both
domestic and export markets can exceed domestic production in years with weaker harvests.

As a result, the sector relies significantly on imported grain, particularly from major exporting
regions in the Black Sea basin. This model has enabled T\"{u}rkiye to create a highly
competitive, export-focused flour industry under stable market conditions. However, it also links
the sector's industrial activity to the availability of imported wheat.

Within the analytical framework introduced above, this configuration corresponds to a relatively
low Biophysical Autonomy Ratio~(BAR). When domestic wheat production is compared with the total
wheat demand generated by domestic consumption and export-oriented processing, BAR values may
fall below unity in years with weaker harvests. In such cases, the effective scale of the milling
industry depends partly on imported grain supplies.

Figure~\ref{fig:Fig_2} illustrates this structural relationship through the evolution
of T\"{u}rkiye's wheat self-sufficiency ratio over the past two decades. The indicator compares
domestic wheat production with the total wheat supply available for domestic use. While
T\"{u}rkiye remains a major wheat producer, the expansion of export-oriented processing has
increased the role of imported grain in sustaining milling activity. However, this apparent
stability conceals an important structural feature: the self-sufficiency ratio does not account
for the additional demand generated by export-oriented processing.

\begin{center}
  \includegraphics[width=\linewidth]{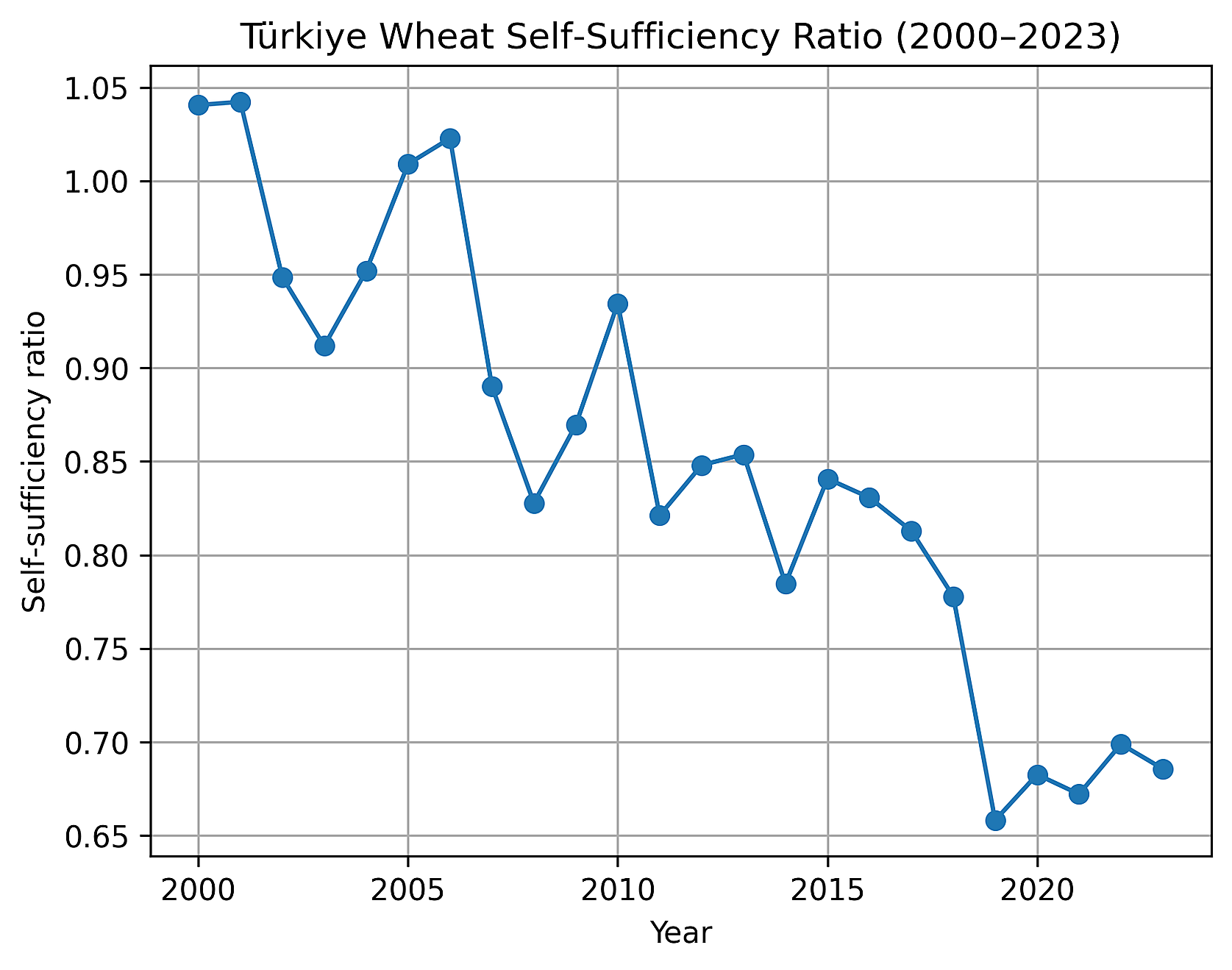}
  \captionof{figure}{Wheat self-sufficiency ratio in T\"{u}rkiye, 2000--2023. The ratio compares domestic wheat production with the total wheat supply available for domestic use
  (production~$+$~imports~$-$~exports). Values below one indicate reliance on imported wheat.
  Data sources: FAOSTAT trade statistics and national production estimates.}
  \label{fig:Fig_2}
\end{center}

\section{Empirical results: Structural patterns in the Biophysical Autonomy Ratio}
\label{sec:results}

The empirical analysis covers a set of countries selected to represent structurally distinct
positions within the global wheat system. Kazakhstan represents a production-surplus exporter
whose domestic agricultural base substantially exceeds processing and consumption demand. Egypt
represents a large import-dependent consumer that relies structurally on global markets for
staple food supply. Germany, Italy, and the Netherlands represent established European processing
economies with varying degrees of domestic production capacity. T\"{u}rkiye is the primary case
under examination. This selection is not exhaustive but is intended to situate T\"{u}rkiye's BAR
trajectory within a range of contrasting system types, allowing structural differences to be
identified through comparison rather than in isolation.

Figure~\ref{fig:Fig_3} reveals clear structural differences in BAR trajectories
across wheat system types.

\begin{center}
  \includegraphics[width=\linewidth]{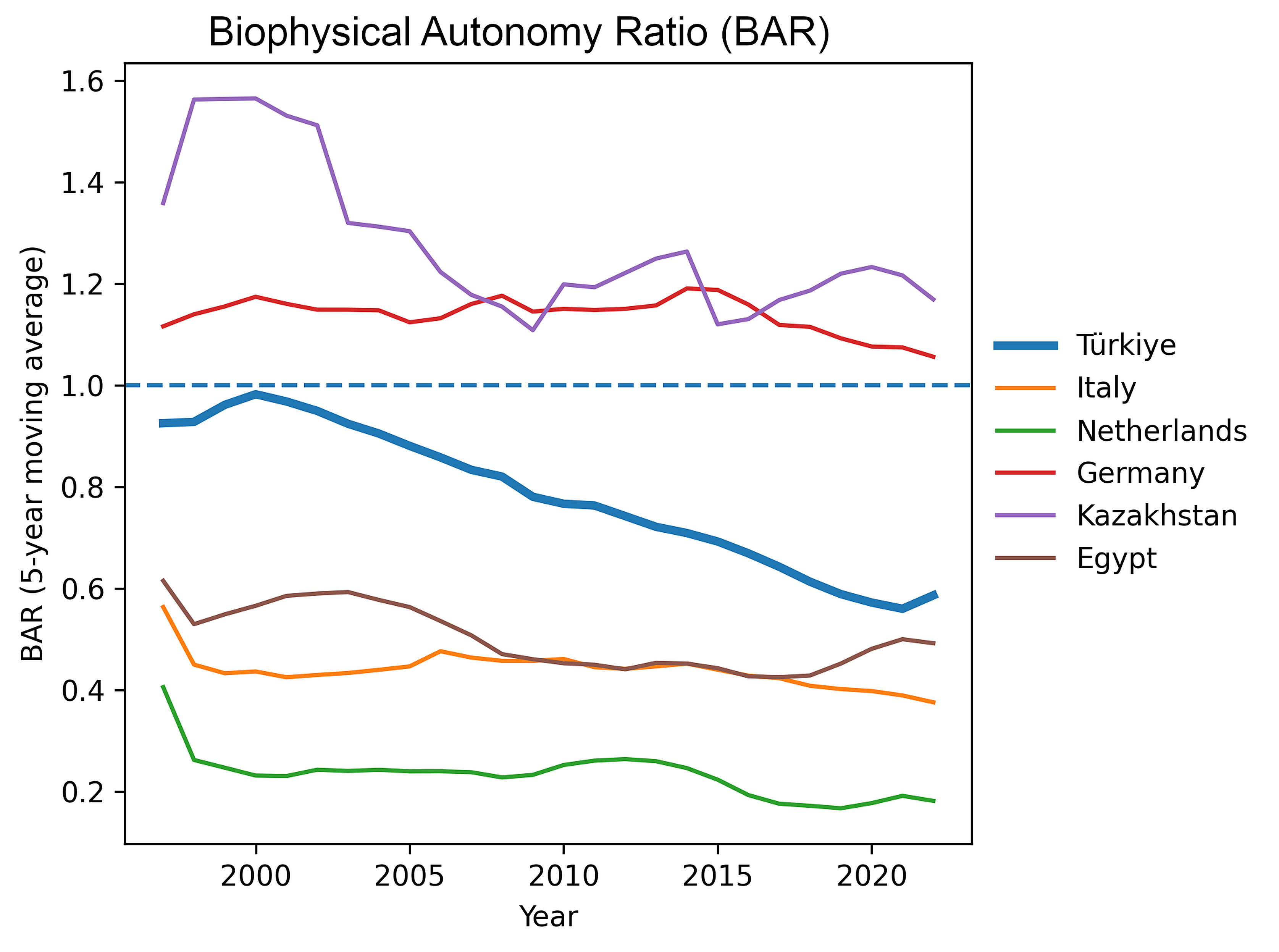}
  \captionof{figure}{Biophysical Autonomy Ratio~(BAR) across wheat system types, 2000--2023. Five-year moving averages are shown to highlight structural trends. The dashed line indicates the
  structural balance threshold~($\mathrm{BAR}=1$), where domestic wheat production is sufficient
  to support both domestic consumption and export-oriented processing demand.
  Production-surplus systems (e.g.\ Kazakhstan) maintain BAR values above unity, while
  import-dependent systems (e.g.\ Egypt) remain consistently below. Processing hubs exhibit more
  heterogeneous patterns: Germany and Italy remain relatively stable, and the Netherlands
  displays structurally low but stable values reflecting import-dependent processing. In contrast,
  T\"{u}rkiye shows a sustained decline in BAR over time, indicating an increasing divergence
  between domestic wheat production and the scale of its export-oriented processing sector. Data
  sources: FAOSTAT production and trade statistics; wheat-equivalent exports calculated using a
  standard extraction coefficient of~0.75.}
  \label{fig:Fig_3}
\end{center}

Table~\ref{tbl:bar_summary} reports the average BAR value for each country during the initial
period (2000--2004) and the most recent period (2019--2023), together with the absolute change
over the observation window. Among the countries examined, T\"{u}rkiye exhibits the largest
decline in BAR, falling from 0.95 to 0.56, which indicates a pronounced structural divergence
between domestic wheat production and the scale of export-oriented processing.

The results reveal clear and consistent differences across system types. Production-surplus
systems, represented by Kazakhstan, maintain BAR values consistently above unity throughout the
observation period. This indicates that domestic production exceeds the combined demands of
domestic consumption and export-oriented processing, reflecting a structural surplus position
within the global wheat system.

In contrast, import-dependent systems, exemplified by Egypt, exhibit BAR values that remain
persistently below one. These systems rely structurally on imported wheat to meet domestic
consumption needs, and their position remains relatively stable over time. This pattern is
consistent with their role within the global food trade as net importers of primary agricultural
commodities.

\begin{table*}[htbp]
\caption{Average Biophysical Autonomy Ratio (BAR) by country for initial (2000--2004) and
recent (2019--2023) periods, and absolute change over the observation window.}
\label{tbl:bar_summary}
\begin{tabular*}{\tblwidth}{@{\extracolsep{\fill}}lcccc@{}}
\toprule
\textbf{Country} & \textbf{BAR (2000--2004)} & \textbf{BAR (2019--2023)} & \textbf{Change} & \textbf{System type} \\
\midrule
Kazakhstan   & 1.51 & 1.22 & $-$0.29 & Production-surplus \\
Germany      & 1.15 & 1.08 & $-$0.07 & Processing hub \\
Italy        & 0.43 & 0.39 & $-$0.04 & Processing hub \\
Netherlands  & 0.24 & 0.19 & $-$0.05 & Processing hub \\
T\"{u}rkiye & 0.95 & 0.56 & $-$0.39 & Processing hub \\
Egypt        & 0.59 & 0.50 & $-$0.09 & Import-dependent \\
\bottomrule
\end{tabular*}
\end{table*}

Processing hubs occupy an intermediate position but display more heterogeneous dynamics.
Countries such as Germany and Italy show relatively stable BAR trajectories, with values
fluctuating around or moderately below unity. The Netherlands, which hosts a significant
processing sector but has limited domestic agricultural production, exhibits consistently low BAR
values. Importantly, these low values reflect a structurally import-dependent processing model
rather than a recent deterioration, and they remain broadly stable over time.

Against this background, T\"{u}rkiye displays a distinct and pronounced trajectory. While BAR
values in the early 2000s are close to unity, indicating a relatively balanced relationship
between domestic production and processing demand, the ratio has sustained a decline over the
subsequent two decades. By the late 2010s and early 2020s, BAR values fell substantially below
one, reaching levels comparable to structurally import-dependent systems.

This decline is not characterised by short-term volatility but by a persistent downward trend,
indicating a structural decoupling between domestic production and processing capacity. Among the
processing hub economies examined, T\"{u}rkiye displays the most pronounced directional shift
over the observation period. While other processing hubs such as Germany and Italy maintain
relatively stable BAR trajectories, and the Netherlands exhibits consistently low but stable
values, T\"{u}rkiye is distinctive in having begun the period in a near-balanced position, with
BAR values approaching unity in the early 2000s, and having undergone sustained structural
deterioration thereafter. This combination of an initially balanced starting point and a
persistent downward trajectory differentiates T\"{u}rkiye from both structurally stable
processing hubs and chronically import-dependent systems and makes it a particularly informative
case for examining how processing-led growth can progressively erode biophysical autonomy.

The observed pattern indicates an increasing divergence between industrial capacity and domestic
biological production. In practical terms, this implies that a growing share of T\"{u}rkiye's
wheat processing activity depends on imported grain inputs rather than domestic production. While
such a configuration can be economically efficient under stable global market conditions, it also
implies greater exposure to external supply dynamics.

According to the analytical framework introduced in this study, the declining BAR trajectory can
be interpreted as a reduction in biophysical autonomy. The Turkish wheat processing system has
become increasingly reliant on external agricultural systems to sustain its current scale of
operation. This structural dependence differentiates T\"{u}rkiye from both surplus-producing
systems and other processing hubs, positioning it as a particularly relevant case for examining
the resilience of globally integrated food processing systems under conditions of increasing
climate variability.

It is worth noting that Italy and the Netherlands have maintained low BAR values throughout the
observation period without apparent disruption to their processing sectors. This suggests that a
low BAR level alone does not indicate imminent instability---as noted in the analytical framework,
these configurations can be economically sustainable under stable trade conditions. The concern
highlighted by T\"{u}rkiye's case is distinct: it is the directional shift from near-autonomy
toward structural import dependence, occurring over a relatively short timeframe, that warrants
attention. A system that has always been import-dependent may have developed adaptive
institutional responses---like diversified supplier networks and established reserve
mechanisms---that a system undergoing rapid structural change may not yet have in place.

Taken together with Figures~\ref{fig:Fig_1} and~\ref{fig:Fig_2}, this result
demonstrates that the apparent stability suggested by conventional indicators conceals a
structural divergence that becomes visible only when export-oriented processing demand is
explicitly incorporated.

\section{Climate risk and supply chain exposure}
\label{sec:climate_risk}

The vulnerability of trade-dependent food processing systems becomes more apparent when
considering the growing climatic instability affecting major global grain-producing regions.
Several of the world's largest wheat-exporting areas---including the Black Sea region, North
America, and parts of Europe---have experienced increasing yield variability associated with
extreme weather events such as droughts, heat waves, and flooding
\citep{Toreti2019, Kornhuber2020, Caparas2021}.

A growing body of research highlights the risk of synchronous production shocks, in which
multiple agricultural regions experience adverse climatic conditions during the same growing
season \citep{Tigchelaar2018, Gaupp2020, Anderson2019}. Such events can significantly reduce
global wheat availability and contribute to rapid price increases in international grain markets.

Under these conditions, exporting countries may respond by restricting exports to stabilise
domestic food prices or protect national food security. Export restrictions played a notable role
in amplifying global food price volatility during previous food crises \citep{Headey2011,
Puma2015}. These policy responses can further tighten global supply and intensify disruptions in
international grain trade.

In terms of the analytical framework introduced earlier, such disruptions reduce the effective
autonomy of processing systems with low BAR values, whose industrial capacity depends on
continued access to imported biological inputs.

For countries with large food processing sectors dependent on imported grain inputs, such
dynamics may create significant exposure to supply chain disruptions. Even when domestic
agricultural production remains stable, reduced access to imported raw materials can constrain
industrial processing activity.

The case of T\"{u}rkiye's flour sector illustrates this structural exposure. As discussed in
Section~\ref{sec:results}, the country's milling industry operates at a scale that frequently
exceeds the domestic wheat production base and therefore relies partly on imported grain,
particularly from the Black Sea region. If major exporting countries experience simultaneous
production shocks or impose export restrictions, the availability of imported wheat could decline
rapidly.

These dynamics suggest that the resilience of globally integrated food processing hubs depends
not only on domestic agricultural production but also on the stability of international supply
chains. As climate-related production variability increases across multiple agricultural regions,
the risk of disruptions affecting trade-dependent processing systems may also grow. From a policy
perspective, this exposure raises questions about the long-term resilience of food systems such
as T\"{u}rkiye's, where large industrial processing capacity coexists with a relatively limited
domestic agricultural production base.

\section{Policy implications}
\label{sec:policy}

The findings of this study carry several implications for food system governance in countries
hosting large export-oriented processing industries.

A central institutional factor in T\"{u}rkiye's case is the Domestic Processing Regime, which
permits the tariff-free importation of wheat for processing and subsequent re-export. This
framework has been instrumental in enabling the growth of T\"{u}rkiye's milling sector and has
generated significant export revenue and industrial employment. At the same time, the analysis
suggests that the regime has contributed to a structural decoupling between processing capacity
and the domestic agricultural production base. As the BAR indicator shows, this decoupling has
deepened progressively over the past two decades. Policymakers may therefore wish to consider
whether the current design of the regime adequately accounts for long-term supply chain exposure,
particularly under conditions of increasing climate variability in major wheat-exporting regions.

This does not necessarily imply a restriction of the regime or a retreat from trade-oriented
industrial policy. Rather, it suggests that resilience considerations could be incorporated
alongside existing competitiveness objectives. For example, BAR-based monitoring could serve as
a diagnostic tool within national food security governance frameworks, helping to identify when
the gap between processing capacity and domestic production has widened to a degree that warrants
policy attention. Similarly, strategic grain reserve requirements calibrated to the volume of
import-dependent processing activity could provide a buffer against short-term supply disruptions
without fundamentally altering the trade framework.

More broadly, the case illustrates a governance challenge that is not unique to T\"{u}rkiye.
Countries that host large food processing industries operating under import-for-re-export regimes
may benefit from periodic assessment of the relationship between industrial capacity and domestic
biological production. Existing food security monitoring frameworks tend to focus on
consumption-side self-sufficiency and may not fully capture the additional demand generated by
export-oriented processing. Incorporating indicators like the BAR into national and international
food security assessments could help close this monitoring gap.

These considerations do not imply that trade-dependent processing models are inherently fragile.
Under stable global market conditions, such models have demonstrated substantial economic
benefits. The concern raised here is more specific: that as climate-related production
variability increases across multiple major exporting regions, the structural exposure of
import-dependent processing hubs may grow in ways that existing governance frameworks are not
well positioned to detect or manage.

\section{Conclusions}
\label{sec:conclusions}

This study examined the relationship between domestic agricultural production and export-oriented
food processing using the Biophysical Autonomy Ratio~(BAR) as an analytical indicator. The
concept provides a simplified framework for evaluating the degree to which industrial food
processing systems depend on domestic biological production capacity or imported agricultural
inputs.

The case of T\"{u}rkiye's wheat and flour sector illustrates how large food processing industries
can operate at a scale that exceeds the domestic agricultural production base. While T\"{u}rkiye
remains a major wheat producer, the expansion of export-oriented milling has created a structural
reliance on imported grain to sustain current levels of industrial activity. Under stable global
market conditions, this model has enabled the development of a highly competitive flour export
industry.

However, increasing climate variability across major grain-producing regions may affect the
reliability of international supply chains. Synchronous production shocks, export restrictions,
or disruptions in global grain trade could limit the availability of imported inputs for
processing industries dependent on external supply.

These dynamics highlight the importance of evaluating the balance between domestic biological
production capacity and industrial processing scale when assessing food system resilience
\citep{Tendall2015}. Analytical indicators such as the BAR may help policymakers identify
structural dependencies and design strategies that strengthen the resilience of globally
integrated food systems while maintaining the benefits of international trade.

In this context, maintaining a minimum level of alignment between industrial processing capacity
and domestic biological production may become a central challenge for food system resilience in
an increasingly unstable climate.

\section*{Declaration of generative AI and AI-assisted technologies in the writing process}

During the preparation of this work, the author used AI-assisted tools (Claude, ChatGPT, and
Gemini) for language editing and structural revision in response to simulated editorial feedback.
All scientific content, data collection, empirical analysis, conceptual framework development,
and conclusions were developed by the author. The author reviewed and takes full responsibility
for all content.

\printcredits

\bibliographystyle{cas-model2-names}
\bibliography{kurnaz2025_BAR}

@ARTICLE{Allan1998,
  author  = {Allan, J.A.},
  title   = {Virtual water: a strategic resource},
  journal = {Ground Water},
  volume  = {36},
  number  = {4},
  year    = {1998},
  pages   = {545--546},
  doi     = {10.1111/j.1745-6584.1998.tb02825.x}
}

@ARTICLE{Anderson2019,
  author  = {Anderson, W.B. and Seager, R. and Baethgen, W. and Cane, M. and You, L.},
  title   = {Synchronous crop failures and climate-forced production variability},
  journal = {Science Advances},
  volume  = {5},
  year    = {2019},
  pages   = {eaaw1976},
  doi     = {10.1126/sciadv.aaw1976}
}

@ARTICLE{Asseng2014,
  author  = {Asseng, S. and Ewert, F. and Martre, P. and R{\"o}tter, R.P. and Lobell, D.B. and
             Cammarano, D. and Kimball, B.A. and Ottman, M.J. and Wall, G.W. and White, J.W. and
             Reynolds, M.P. and Alderman, P.D. and Prasad, P.V.V. and Aggarwal, P.K. and
             Anothai, J. and Basso, B. and Biernath, C. and Challinor, A.J. and De Sanctis, G. and
             Doltra, J. and Fereres, E. and Garcia-Vila, M. and Gayler, S. and Hoogenboom, G. and
             Hunt, L.A. and Izaurralde, R.C. and Jabloun, M. and Jones, C.D. and Kersebaum, K.C. and
             Koehler, A.-K. and M{\"u}ller, C. and Naresh Kumar, S. and Nendel, C. and O'Leary, G. and
             Olesen, J.E. and Palosuo, T. and Priesack, E. and Eyshi Rezaei, E. and Ruane, A.C. and
             Semenov, M.A. and Shcherbak, I. and St{\"o}ckle, C. and Stratonovitch, P. and
             Streck, T. and Supit, I. and Tao, F. and Thorburn, P.J. and Waha, K. and Wang, E. and
             Wallach, D. and Wolf, J. and Zhao, Z. and Zhu, Y.},
  title   = {Rising temperatures reduce global wheat production},
  journal = {Nature Climate Change},
  volume  = {5},
  year    = {2014},
  pages   = {143--147},
  doi     = {10.1038/nclimate2470}
}

@ARTICLE{Biess2024,
  author  = {Biess, B. and Gudmundsson, L. and Windisch, M.G. and Seneviratne, S.I.},
  title   = {Future changes in spatially compounding hot, wet or dry events and their implications
             for the world's breadbasket regions},
  journal = {Environmental Research Letters},
  volume  = {19},
  year    = {2024},
  pages   = {064011},
  doi     = {10.1088/1748-9326/ad4619}
}

@ARTICLE{BrenDAmour2016,
  author  = {Bren d'Amour, C. and Wenz, L. and Kalkuhl, M. and Steckel, J.C. and Creutzig, F.},
  title   = {Teleconnected food supply shocks},
  journal = {Environmental Research Letters},
  volume  = {11},
  number  = {3},
  year    = {2016},
  pages   = {035007},
  doi     = {10.1088/1748-9326/11/3/035007}
}

@ARTICLE{Caparas2021,
  author  = {Caparas, M. and Zobel, Z. and Castanho, A.D.A. and Schwalm, C.R.},
  title   = {Increasing risks of crop failure and water scarcity in global breadbaskets by 2030},
  journal = {Environmental Research Letters},
  volume  = {16},
  year    = {2021},
  pages   = {104013},
  doi     = {10.1088/1748-9326/ac22c1}
}

@ARTICLE{Chatzopoulos2021,
  author  = {Chatzopoulos, T. and P{\'e}rez Dom{\'i}nguez, I. and Toreti, A. and Aden{\"a}uer, M. and Zampieri, M.},
  title   = {Potential impacts of concurrent and recurrent climate extremes on the global food
             system by 2030},
  journal = {Environmental Research Letters},
  volume  = {16},
  year    = {2021},
  pages   = {124021},
  doi     = {10.1088/1748-9326/ac343b}
}

@ARTICLE{Clapp2017,
  author  = {Clapp, J.},
  title   = {Responsibility to the rescue? {G}overning private financial investment in global agriculture},
  journal = {Agriculture and Human Values},
  volume  = {34},
  year    = {2017},
  pages   = {223--235},
  doi     = {10.1007/s10460-015-9678-8}
}

@ARTICLE{Cottrell2019,
  author  = {Cottrell, R.S. and Nash, K.L. and Halpern, B.S. and Remenyi, T.A. and Corney, S.P. and
             Fleming, A. and Fulton, E.A. and Hornborg, S. and Johne, A. and Watson, R.A. and
             Blanchard, J.L.},
  title   = {Food production shocks across land and sea},
  journal = {Nature Sustainability},
  volume  = {2},
  year    = {2019},
  pages   = {130--137},
  doi     = {10.1038/s41893-018-0210-1}
}

@ARTICLE{Deteix2024,
  author  = {Deteix, L. and Salou, T. and Loiseau, E.},
  title   = {Quantifying food consumption supply risk: an analysis across countries and agricultural products},
  journal = {Global Food Security},
  volume  = {41},
  year    = {2024},
  pages   = {100764},
  doi     = {10.1016/j.gfs.2024.100764}
}

@ARTICLE{Fader2011,
  author  = {Fader, M. and Gerten, D. and Thammer, M. and Heinke, J. and Lotze-Campen, H. and
             Lucht, W. and Cramer, W.},
  title   = {Internal and external green-blue agricultural water footprints of nations, and related
             water and land savings through trade},
  journal = {Hydrology and Earth System Sciences},
  volume  = {15},
  year    = {2011},
  pages   = {1641--1660},
  doi     = {10.5194/hess-15-1641-2011}
}

@ARTICLE{Francis2012,
  author  = {Francis, J.A. and Vavrus, S.J.},
  title   = {Evidence linking {A}rctic amplification to extreme weather in mid-latitudes},
  journal = {Geophysical Research Letters},
  volume  = {39},
  year    = {2012},
  pages   = {L06801},
  doi     = {10.1029/2012GL051000}
}

@ARTICLE{Gaupp2019,
  author  = {Gaupp, F. and Hall, J. and Mitchell, D. and Dadson, S.},
  title   = {Increasing risks of multiple breadbasket failure under 1.5 and 2{$^\circ$C} global warming},
  journal = {Agricultural Systems},
  volume  = {175},
  year    = {2019},
  pages   = {34--45},
  doi     = {10.1016/j.agsy.2019.05.010}
}

@ARTICLE{Gaupp2020,
  author  = {Gaupp, F. and Hall, J. and Hochrainer-Stigler, S. and Dadson, S.},
  title   = {Changing risks of simultaneous global breadbasket failure},
  journal = {Nature Climate Change},
  volume  = {10},
  year    = {2020},
  pages   = {54--57},
  doi     = {10.1038/s41558-019-0600-z}
}

@ARTICLE{Hasegawa2022,
  author  = {Hasegawa, T. and Wakatsuki, H. and Nelson, G.C.},
  title   = {Evidence for and projection of multi-breadbasket failure caused by climate change},
  journal = {Current Opinion in Environmental Sustainability},
  volume  = {58},
  year    = {2022},
  pages   = {101217},
  doi     = {10.1016/j.cosust.2022.101217}
}

@ARTICLE{Headey2011,
  author  = {Headey, D.},
  title   = {Rethinking the global food crisis: the role of trade shocks},
  journal = {Food Policy},
  volume  = {36},
  year    = {2011},
  pages   = {136--146},
  doi     = {10.1016/j.foodpol.2010.10.003}
}

@ARTICLE{Heino2020,
  author  = {Heino, M. and Guillaume, J.H.A. and M{\"u}ller, C. and Iizumi, T. and Kummu, M.},
  title   = {A multi-model analysis of teleconnected crop yield variability in a range of cropping systems},
  journal = {Earth System Dynamics},
  volume  = {11},
  year    = {2020},
  pages   = {113--128},
  doi     = {10.5194/esd-11-113-2020}
}

@ARTICLE{Hoekstra2005,
  author  = {Hoekstra, A.Y. and Hung, P.Q.},
  title   = {Globalisation of water resources: international virtual water flows in relation to
             crop trade},
  journal = {Global Environmental Change},
  volume  = {15},
  year    = {2005},
  pages   = {45--56},
  doi     = {10.1016/j.gloenvcha.2004.06.004}
}

@ARTICLE{Keys2025,
  author  = {Keys, P.W. and Barnes, E.A. and Diffenbaugh, N.S. and Hertel, T.W. and
             Baldos, U.L.C. and Hedlund, J.},
  title   = {Exposure to compound climate hazards transmitted via global agricultural trade networks},
  journal = {Environmental Research Letters},
  volume  = {20},
  year    = {2025},
  pages   = {044039},
  doi     = {10.1088/1748-9326/adb86a}
}

@ARTICLE{Kornhuber2020,
  author  = {Kornhuber, K. and Coumou, D. and Vogel, E. and Lesk, C. and Donges, J.F. and
             Lehmann, J. and Horton, R.M.},
  title   = {Amplified {R}ossby waves enhance risk of concurrent heatwaves in major breadbasket regions},
  journal = {Nature Climate Change},
  volume  = {10},
  year    = {2020},
  pages   = {48--53},
  doi     = {10.1038/s41558-019-0637-z}
}

@ARTICLE{Lesk2016,
  author  = {Lesk, C. and Rowhani, P. and Ramankutty, N.},
  title   = {Influence of extreme weather disasters on global crop production},
  journal = {Nature},
  volume  = {529},
  year    = {2016},
  pages   = {84--87},
  doi     = {10.1038/nature16467}
}

@ARTICLE{Lobell2011,
  author  = {Lobell, D.B. and B{\"a}nziger, M. and Magorokosho, C. and Vivek, B.},
  title   = {Nonlinear heat effects on {A}frican maize as evidenced by historical yield trials},
  journal = {Nature Climate Change},
  volume  = {1},
  year    = {2011},
  pages   = {42--45},
  doi     = {10.1038/nclimate1043}
}

@ARTICLE{Mehrabi2019,
  author  = {Mehrabi, Z. and Ramankutty, N.},
  title   = {Synchronized failure of global crop production},
  journal = {Nature Ecology \& Evolution},
  volume  = {3},
  year    = {2019},
  pages   = {780--786},
  doi     = {10.1038/s41559-019-0862-x}
}

@ARTICLE{Puma2015,
  author  = {Puma, M.J. and Bose, S. and Chon, S.Y. and Cook, B.I.},
  title   = {Assessing the evolving fragility of the global food system},
  journal = {Environmental Research Letters},
  volume  = {10},
  year    = {2015},
  pages   = {024007},
  doi     = {10.1088/1748-9326/10/2/024007}
}

@ARTICLE{Qi2022,
  author  = {Qi, W. and Feng, L. and Yang, H. and Liu, J.},
  title   = {Increasing concurrent drought probability in global main crop production countries},
  journal = {Geophysical Research Letters},
  volume  = {49},
  year    = {2022},
  pages   = {e2021GL097060},
  doi     = {10.1029/2021GL097060}
}

@ARTICLE{Raymond2020,
  author  = {Raymond, C. and Horton, R.M. and Zscheischler, J. and Martius, O. and
             AghaKouchak, A. and Balch, J. and Bowen, S.G. and Camargo, S.J. and Hess, J. and
             Kornhuber, K. and Oppenheimer, M. and Ruane, A.C. and Wahl, T. and White, K.},
  title   = {Understanding and managing connected extreme events},
  journal = {Nature Climate Change},
  volume  = {10},
  year    = {2020},
  pages   = {611--621},
  doi     = {10.1038/s41558-020-0790-4}
}

@ARTICLE{Rulli2013,
  author  = {Rulli, M.C. and Saviori, A. and D'Odorico, P.},
  title   = {Global land and water grabbing},
  journal = {Proceedings of the National Academy of Sciences},
  volume  = {110},
  number  = {3},
  year    = {2013},
  pages   = {892--897},
  doi     = {10.1073/pnas.1213163110}
}

@ARTICLE{Schauberger2017,
  author  = {Schauberger, B. and Archontoulis, S. and Arneth, A. and Balkovic, J. and Ciais, P. and
             Deryng, D. and Elliott, J. and Folberth, C. and Khabarov, N. and M{\"u}ller, C. and
             Pugh, T.A.M. and Rolinski, S. and Schaphoff, S. and Schmid, E. and Wang, X. and
             Schlenker, W. and Frieler, K.},
  title   = {Consistent negative response of {US} crops to high temperatures in observations and
             crop models},
  journal = {Nature Communications},
  volume  = {8},
  year    = {2017},
  pages   = {13931},
  doi     = {10.1038/ncomms13931}
}

@ARTICLE{Tendall2015,
  author  = {Tendall, D.M. and Joerin, J. and Kopainsky, B. and Edwards, P. and Shreck, A. and
             Le, Q.B. and Kr{\"u}tli, P. and Grant, M. and Six, J.},
  title   = {Food system resilience: defining the concept},
  journal = {Global Food Security},
  volume  = {6},
  year    = {2015},
  pages   = {17--23},
  doi     = {10.1016/j.gfs.2015.08.001}
}

@ARTICLE{Tigchelaar2018,
  author  = {Tigchelaar, M. and Battisti, D.S. and Naylor, R.L. and Ray, D.K.},
  title   = {Future warming increases probability of globally synchronized maize production shocks},
  journal = {Proceedings of the National Academy of Sciences USA},
  volume  = {115},
  year    = {2018},
  pages   = {6644--6649},
  doi     = {10.1073/pnas.1718031115}
}

@ARTICLE{Toreti2019,
  author  = {Toreti, A. and Cronie, O. and Zampieri, M.},
  title   = {Concurrent climate extremes in the key wheat producing regions of the world},
  journal = {Scientific Reports},
  volume  = {9},
  year    = {2019},
  pages   = {5493},
  doi     = {10.1038/s41598-019-41932-5}
}

@ARTICLE{Zhao2017,
  author  = {Zhao, C. and Liu, B. and Piao, S. and Wang, X. and Lobell, D.B. and Huang, Y. and
             Huang, M. and Yao, Y. and Bassu, S. and Ciais, P. and Durand, J.-L. and Elliott, J. and
             Ewert, F. and Janssens, I.A. and Li, T. and Lin, E. and Liu, Q. and Martre, P. and
             M{\"u}ller, C. and Peng, S. and Pe{\~n}uelas, J. and Ruane, A.C. and Wallach, D. and
             Wang, T. and Wu, D. and Liu, Z. and Zhu, Y. and Zhu, Z. and Asseng, S.},
  title   = {Temperature increase reduces global yields of major crops in four independent estimates},
  journal = {Proceedings of the National Academy of Sciences},
  volume  = {114},
  year    = {2017},
  pages   = {9326--9331},
  doi     = {10.1073/pnas.1701762114}
}

\end{document}